\documentclass[journal,a4paper,10pt]{IEEEtran}
\usepackage{amsmath,amssymb,amsfonts} 
\usepackage{graphicx}                 
\usepackage{fontenc}
\usepackage{multirow} 
\usepackage{theorem}
\usepackage{array}
\usepackage{flushend}

\usepackage[footnotesize]{caption}
\usepackage{color,cite}
\usepackage{algorithmic}
\usepackage{algorithm}
\usepackage{epstopdf}
\usepackage{mathrsfs}
\usepackage[english]{babel}

\makeatother

\ifodd 1
\newcommand{\comg}[1]{{\color{black} #1}} 
\newcommand{\com}[1]{\textbf{\color{red} #1}}
\newcommand{\response}[1]{\textbf{\color{magenta} (RESPONSE: #1)}} 
\else

\newcommand{\com}[1]{}
\newcommand{\comg}[1]{}
\newcommand{\response}[1]{}

\fi

\begin{document}
\title{\comg{Guaranteeing QoS using Unlicensed TV White Spaces for Smart Grid Applications}}
\author{Naveed Ul Hassan,~\IEEEmembership{Senior Member,~IEEE,}~Wayes Tushar,~\IEEEmembership{Member,~IEEE,}  ~Chau~Yuen,~\IEEEmembership{Senior Member,~IEEE},~See Gim Kerk~\IEEEmembership{Member,~IEEE,} and Ser Wah Oh,~\IEEEmembership{Senior Member,~IEEE}
\thanks{This work is supported in part by the Lahore University of Management Sciences (LUMS), in part by the Singapore grants NRF2012EWT-EIRP002-045 and NRF2015ENC-GBICRD001-028, in part by A*Star project 1420200043,  and in part by the SUTD-MIT International Design Centre (IDC:idc.sutd.edu.sg).}
\thanks{N. U. Hassan is with the Electrical Engineering Department at Lahore University of Management Sciences (LUMS), Lahore, Pakistan 54792 (Email: naveed.hassan@lums.edu.pk).}
\thanks{W. Tushar and C. Yuen are with the Singapore University of Technology and Design (SUTD), 8 Somapah Road, Singapore 487372 (Email: \{wayes\_tushar, yuenchau\}@sutd.edu.sg).}
\thanks{S. G. Kerk is with the Power Automation, Singapore (Email: sgkerk@ieee.org)}
\thanks{S. W. Oh is with the Whizpace Pte Ltd, Singapore (Email: swoh@i2r.a-star.edu.sg)}
}
\IEEEoverridecommandlockouts
\maketitle
\begin{abstract}
In this paper, we consider the utilization of TV White Spaces (TVWS) by small Cognitive Radio (CR) network operators to support the communication needs of various smart grid applications. We first propose a multi-tier communication network architecture for smart metering applications in dense urban environments. Our measurement campaign, without any competition from other CR operators, reveals that the communication architecture can achieve more than 1Mbps data rates using the free unlicensed TVWS spectrum. However, anticipating stiff competition for the unlicensed TVWS spectrum among CR operators and to support smart grid applications with stringent Quality of Service (QoS) requirements, we further exploit the novel idea of high priority channels (HPC) that the CR operator can temporarily lease by paying a small fee. This poses several new challenges for the CR operators, such as, their economic viability while providing QoS guarantees. We develop a real-time decision support framework with several adjustable parameters for the CR operators that enables them to tradeoff HPC leasing cost and QoS. The developed algorithms are simple rules that provide significant opportunities to the CR operators to maintain a balance between spectrum cost and QoS depending on dynamic spectrum availability and smart grid application requirements.     

\end{abstract}
 \setcounter{page}{1}
\section{Introduction}\label{sec:introduction}
Smart grid is a green technology that allows the integration of renewable energy sources and demand response management. Internet of Things (IoT) and Machine-to-Machine (M2M) communications are the major drivers of smart grid deployment and applications, particularly in residential and commercial buildings. IoT is a broad futuristic vision where a network of ``things" such as sensors, objects and nodes is formed to connect and unite physical and digital worlds~\cite{Lin:2016,sanjana-IoT:2016}. The ``things" are often called ``machines" when they sense, collect and communicate useful data with minimal human intervention. M2M communication is used for automated data transmission between various machines. Currently, M2M connections account for 2.8\% of the total worldwide mobile connections and nearly 428 mobile operators offer M2M services across 187 countries \cite{GSMA:2014}. It is expected that by the year 2022, there will be around 18 billion M2M connections. Addressing the ever increasing data rate demands of dense M2M networks in smart grids requires novel network architectures and an efficient utilization of communication resources~\cite{Kim:2014}.

It is needless to point out the scarcity of wireless spectrum and the usefulness of Cognitive Radio (CR) technology that allows the opportunistic reuse of wireless spectrum. In recent years, due to the transition from analog to digital TV broadcasts, some spectrum in the very high frequency (VHF) and ultra high frequency (UHF) bands are also becoming available. This spectrum is commonly known as the TV white spaces (TVWS)~\cite{Bogucka:2012}. Several countries have already laid out regulatory frameworks for TVWS utilization, e.g., the Federal Communications Commission (FCC) in USA~\cite{fccc}, the Office of Communications (OFCOM) in UK~\cite{of_com} and Info-Communications Development Authority (IDA) in Singapore~\cite{sing_ida}. There is a general consensus among the spectrum regulators to provide an unlicensed access to TVWS. Given the attractive propagation characteristics in the VHF and UHF bands and relatively larger frequency blocks, there are also some efforts to standardize the use of TVWS spectrum. For example, the IEEE 802.11af, also sometimes referred to as `White-Fi' or `Super Wi-Fi' adds the TVWS bands to the IEEE 802.11 family of specifications. TVWS has been touted as an untapped resource that has the potential to unleash a myriad of new M2M applications and services. Moreover, unlike traditional licensed wireless operators, free availability of TVWS spectrum provides a great opportunity for many new small and virtual CR operators to setup their networks for smart grid applications at very low cost. In this paper, any operator using free unlicensed spectrum bands without owning any other wireless spectrum bands is termed as a small and virtual CR operator.  

To ensure non-interfering spectrum access, a CR operator can acquire spectrum availability information either through i) spectrum sensing or ii) by contacting a geo-location database \cite{rev_ref}. In spectrum sensing technique, CR operator is responsible for spectrum sensing and managing its transmissions to avoid unnecessary interference. In geo-location database method, CR operator obtains spectrum information from a service provider that maintains a complete database of spectrum availability at different locations by employing sophisticated spectrum sensing techniques. Geo-location database method has some additional signaling overhead for the CR operators. However, this technique is preferred by the spectrum regulators to provide non-interfering spectrum access to competing CR operators in unlicensed TVWS bands~\cite{sing_ida}, \cite{tvws1}. 

Many novel smart grid applications, e.g., demand response management, energy monitoring for shared facility (e.g. lift, corridor light), environmental sensing, etc., would require higher data rates with more stringent Quality of Service (QoS) requirements. To cater for these situations, already there are proposals to assign certain channels in the TVWS spectrum as high priority channels (HPCs)~\cite{sing_ida}. The HPCs could be leased, temporarily, by the interested CR operators for exclusive short term usage by paying a small license fee to the geo-location database provider. Maintaining economic viability would then require appropriate HPC purchase decisions by the CR operator and in a highly dynamic environment, with varying spectrum availability, channel conditions and incoming data traffic, balancing HPC leasing cost and QoS guarantees could become quite challenging. 

In this paper, we study the performance of a communication network architecture for M2M applications in smart grids using TVWS. Following the approach in \cite{crn3}, a multi-tier communication network is established. In the proposed architecture, building area network (BAN) is based on power line communications (PLC), neighborhood area network (NAN) is based on TVWS, and wide area network (WAN) relies on cable or fiber-optics. We implement the communication network in large residential complexes in Singapore that could be used to transmit the aggregated data of several hundred smart meters using TVWS spectrum to the remote servers of the network operator. We launched a measurement campaign in the dense urban environment of Singapore to determine the achieved data rates. The lowest data rates that we obtained using TVWS spectrum exceeded 1Mbps, which are sufficient for several smart grid applications. It is important to note that currently in Singapore, TVWS is in trial phase and we did not face any spectrum competition from other CR operators. 

Considering future growth and interest in smart grid applications, we however, anticipate an overcrowding of TVWS, amid stiff competition among several CR operators. CR operators would be required to lease HPCs to guarantee QoS demands of numerous smart grid applications\footnote{It is important to note that in this scenario our work focuses on the TVWS spectrum purchase and QoS issues faced by the CR operators, regardless of the nature of underlying protocol (IEEE 802.11af, etc.)}. In this context, we develop a simple threshold based real-time decision support mechanism based on Lyapunov optimization framework to help the CR operator in making the HPC purchase decisions. The proposed mechanism captures the tradeoff between the HPC leasing cost and transmission delay and is named as `Dynamic-CD-LYAPUNOV' scheme. We then consider data quality as another QoS dimension. This consideration is based on the fact that in some smart grid applications, data quality, measured in terms of the size of data unit (packets, frames), could be reduced, without any significant impact on the overall performance. For example, depending on spectrum availability, the least significant bits of smart meter readings or environmental sensor readings could be dropped (exploiting redundancy in the readings). Considering two QoS dimensions, we study tradeoff between spectrum cost, transmission delay and data quality and develop a `Dynamic-CD-QUALITY' scheme. Simulation results indicate that the schemes developed in this work provide significant opportunities to the CR operators to maintain a balance between spectrum cost and QoS for several emerging smart grid applications. 

To summarize, there are three main contributions of this work. First, we propose a communication network architecture for CR operators that exploit TVWS spectrum and present the results of a measurement campaign in dense urban environment to determine the achievable data rates for smart grid applications. Second, we develop a threshold based real-time decision rule for the CR operator to make HPC leasing decisions and adjustable parameters to tradeoff spectrum leasing cost and queuing delay. Finally, we propose an algorithm that exploits data quality along with the transmission delay to reduce the HPC leasing cost. These novel ideas and the resulting optimization problems are not considered in any prior work on smart grids.  
 
The rest of the paper is organized as follows. In Section~\ref{sec:network}, we describe the proposed communication network architecture and present the achievable data rates in dense urban environment using TVWS spectrum. Dynamic-CD-LYAPUNOV scheme is developed in Section~\ref{sec:Qos}, while Dynamic-CD-QUALITY scheme is presented in Section~\ref{sec:Qual}. In Section~\ref{sec:sims}, we present the simulation results and conclude the paper in Section~\ref{sec:conc}.

\section{Proposed Smart Grid Communication Network Architecture and the Results of a Measurement Campaign} 
\label{sec:network}
In this section, we discuss a generic communication network architecture, exploiting TVWS, for dense M2M communication networks supporting smart grid applications, such as, demand response management, home automation, data analysis, etc. Based on the proposed architecture, we also present the achievable data rates in dense urban environment using TVWS spectrum. 

\subsection{Proposed Smart Grid Communication Network Architecture}
The proposed architecture can be readily deployed in any residential, commercial or industrial complex, having numerous buildings and a large number of machine type devices. The choice of an appropriate communication technology at the building, neighborhood and wide area network levels that could leverage existing infrastructure or freely available resources, while satisfying the data rate requirements of the targeted application is essential for a cost effective solution. 

Usually each machine (sensor, smart meter, etc.,) inside a building generates or collects some data, which is transmitted by establishing a BAN. To create a BAN, we propose wired access i.e., PLC, which re-uses existing power lines for communication. Unlike Wi-Fi and other wireless personal area technologies (e.g., Zigbee, Bluetooth), PLC is more secure, does not create interference and provides superior performance across thick walls. PLC also simplifies the network architecture and management of machines since power lines are readily available in most parts of the buildings. The data collected from several machines is further aggregated at the data concentrator that may be located on the rooftop or some other convenient location in the building. 

The data concentrator, deployed by the CR operator, connects to the remote base station (BS) and establishes a NAN. There are numerous possible communication technologies to build a NAN e.g., through wired access (cable, fiber-optics) or wireless technology (TVWS, licensed spectrum). Connecting every building in the neighborhood via a wired access is less flexible and costly because such wired infrastructure does not exist. Building a NAN on TVWS spectrum, on the other hand, provides operational flexibility to the CR operators because they can easily scale up or down their networks by adding or removing BSs. We, therefore, consider a TVWS based NAN in this paper. CR operator can obtain TVWS spectrum information either by contacting the geo-location database or through spectrum sensing\footnote{The architecture and the decision support mechanism proposed in this paper can always be used regardless of the method adopted by the CR operator to obtain TVWS spectrum information.}. Once the aggregated data reaches the BS, it is further transmitted to the remote servers of the network operator using cable or fiber-optics.   

\begin{figure}
\centering
\includegraphics[width=\columnwidth]{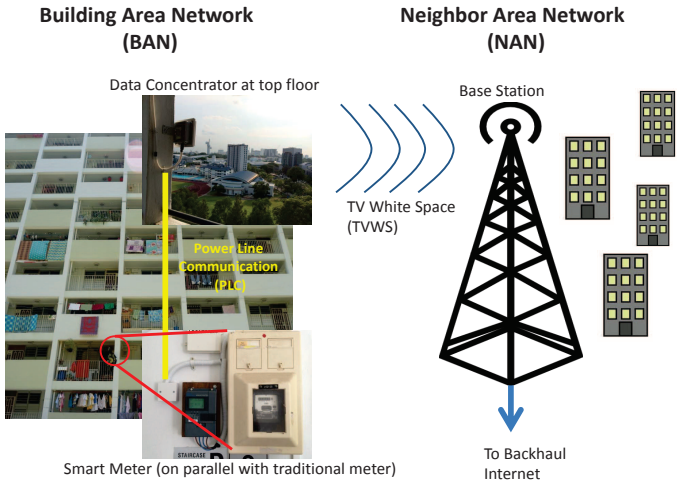}
\caption{Testbed for smart metering application at a Residential Block}
\label{fig:fig1}
\end{figure}

Based on the proposed communication network architecture, we have built a testbed for smart metering applications. Communication requirements of a smart metering application in a dense urban environment, like Singapore, with a large number of high-rise buildings, comprising of several hundred apartments, each equipped with an individual smart meter could be huge (several Mbps). Our proposed communication network architecture to cater for the communication needs of such an application is shown in Figure~\ref{fig:fig1}.

The performance and throughput of PLC employed in BAN and cable or fiber-optics in WAN are well known in the literature. The focus of our measurement campaign exclusively pertains to the achieved data rates on TVWS at various locations in the dense urban environment. Commercial-off-the-shelf, TVWS transceivers are not readily available in the market. Therefore, in this work, we have used our own TVWS transceivers. The specifications of the TVWS based NAN, established in our testbed are given in Table I. The interference among the data concentrators is avoided by adopting a time division multiple access (TDMA) schedule. In TDMA schedule, each data concentrator utilizes all the available wireless spectrum only during its designated time slot, while the remaining data concentrators remain silent. This scheduling is adopted for its simplicity, practicality and less feedback and signaling overheads. 
\begin{table}
\centering
\caption{Specifications of TVWS and data concentrator.}
\includegraphics[width=\columnwidth]{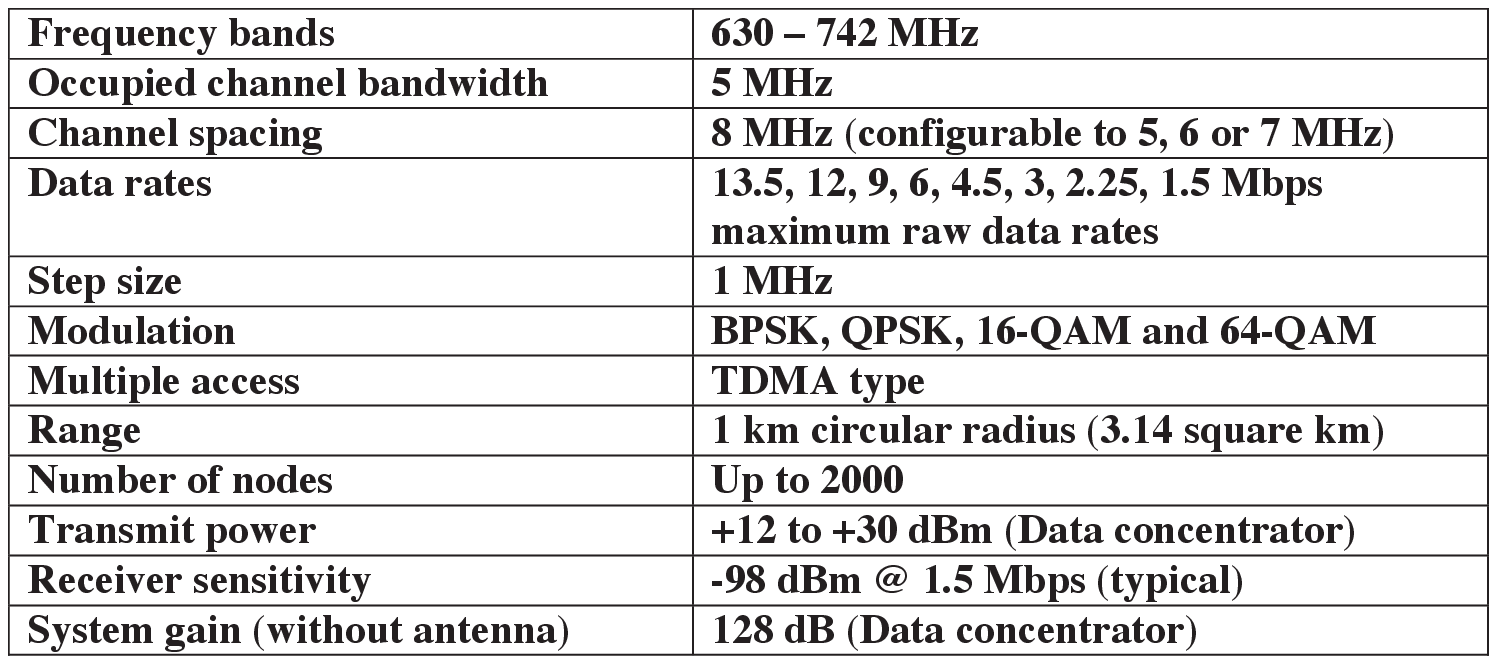}
\label{table:1}
\end{table}

\subsection{Test Results of Measurement Campaign in an Urban Environment}
\begin{figure*}
\centering
\includegraphics[width=0.75\textwidth,height=.45\textheight]{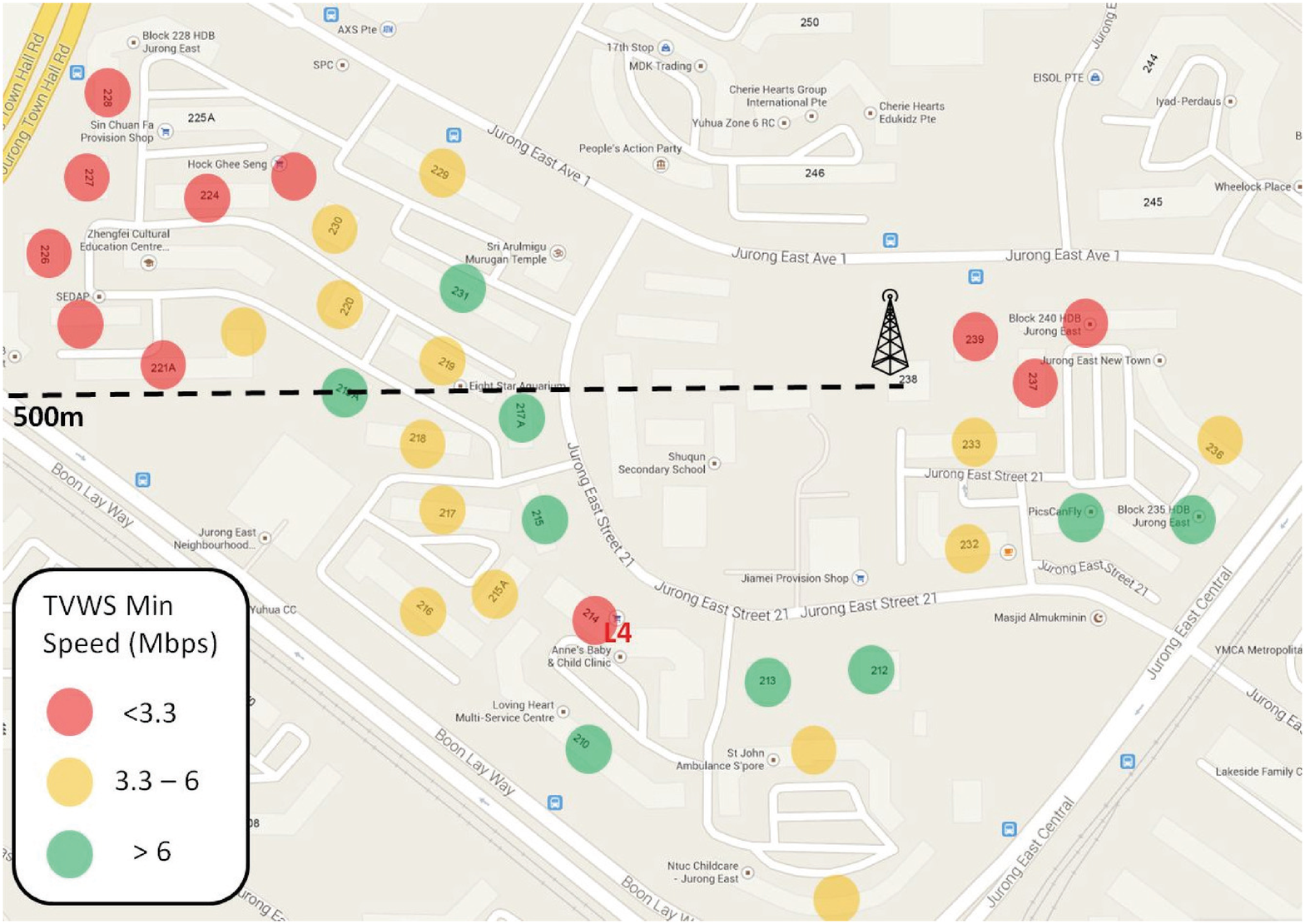}
\caption{Results of Measurement Campaign around Jurong East Area, Singapore using TVWS spectrum.}
\label{fig:figr}
\end{figure*}

We performed a measurement campaign in the Jurong East area of Singapore as shown in Fig.~\ref{fig:figr}. As shown in Figure~\ref{fig:figr}, we installed a BS at Level 25 of Block 238. We then measured the TVWS connectivity at several blocks of buildings that are scattered within 500m radius of the BS. Each block in our study is a high-rise multi-level building, with several residential units at each level (the number of levels in each building varies from 10 to 40). The data concentrator in each block is installed at a different level depending on the site suitability. In our measurement campaign, there are no other competing CR operators for the unlicensed TVWS spectrum and we are always able to get the required spectrum for our application. 

As expected, the achieved data rates vary depending on the distance from the BS, direction of the antenna and the building height.
Physical constraints on site availability for data concentrator antenna direction, lower available height at which data concentrator could be installed and the distance of the building from BS are few of the reasons that often resulted in lower data rates. In Figure~\ref{fig:figr}, all such bad locations are marked as red colored nodes. However, even at these bad locations, we achieved a throughput of at least 1Mbps. These results clearly demonstrate the effectiveness of TVWS to support smart grid applications. However, the success of TVWS in providing high data rates (several Mbps) primarily depends on the ready availability of free TVWS spectrum. In future, CR operators are expected to lease HPCs to fulfill the data rate requirements of some smart grid applications as the competition for free TVWS spectrum will grow among the operators. Novel tradeoffs and decision support mechanisms are required to be developed in case of TVWS spectrum congestion to minimize the operational cost of CR operators and provide acceptable QoS guarantees.

\section{Dynamic-CD-LYAPUNOV Scheme: Mechanism to tradeoff HPC leasing cost and transmission delays}
\label{sec:Qos}
In this section, we develop our Dynamic-CD-LYAPUNOV scheme. This is a Lyapunov optimization based decision making process for the CR operators to tradeoff HPC leasing cost and the transmission delays. The technique leads to a simple threshold based HPC purchase rule that can be implemented in real-time by the CR operators.

\subsection{System Model and Problem Formulation}
In our communication network architecture, TVWS is used to establish a NAN that comprises of $K$ buildings, each equipped with one data concentrator. The data packets generated by the smart meters or the environmental sensors in each building are aggregated and held in a queue maintained at the data concentrator. We assume a time-slotted transmission system and TDMA scheduling, where each transmission time interval is divided into $K$ sub-slots. Each data concentrator transmits its queued packets in its own reserved time slot. 
Let, $Q_k(t)$ denotes the queue length at data concentrator $k$ at time $t$. The queue length depends on the packet arrival and departure rates that are respectively denoted by $A_k(t)$ and $R_k(t)$. We assume an upper bound on the packet arrival and the departure rates in the queue. We also assume a dynamic spectrum leasing price, where $c(t)$ denotes the cost of transmitting one data unit (packets, group of packets) using the leased HPCs in time slot $t$. In our model, we assume that all the data concentrators report their queue lengths to the remote BS where HPC purchase decisions are made and accordingly communicated to the data concentrators. It is important to note that the developed Dynamic-CD-LYAPUNOV scheme can also be individually implemented by each data concentrator.

In a queuing based system model, QoS can be linked to the amount of time that is spent by a packet in the queue before its transmission. We also call it the transmission delay or the queuing delay. Higher queuing delay means lower QoS and vice-verse. We consider Lyapunov optimization technique that leads to a real-time threshold based decision mechanism, while ensuring the stability of the queues (i.e., the queue lengths remain finite). This method also bounds the worst case queuing delay and provides some adjustable parameters that allows tradeoff between HPC purchase decisions and transmission delay for different smart grid applications \cite{Neely1,Neely2}. 

\subsection{Real-Time Dynamic-CD-LYAPUNOV Scheme}
A typical data queue, $Q_k(t)$, has no awareness of queuing delays. In order to keep track of queuing delays, BS defines a delay aware virtual queue for each data concentrator. Let $Z_k(t)$ denote the length of the delay aware virtual queue of data concentrator $k$ at time $t$. The service process of $Z_k(t)$ and $Q_k(t)$ are the same i.e., $R_k(t)$. However, the virtual queue, $Z_k(t)$, increments the queue length by an amount $\epsilon>0$, whenever the actual queue is non-empty. This increment indicates the presence of packets in the queue that are not serviced and creates delay awareness. 

Let, $Y_k(t)=Q_k(t)+Z_k(t)$, denotes the sum of actual and virtual queue lengths of data concentrator $k$ at time $t$. Let, $D_k(t) \in \{0,1\}$, denotes a binary decision variable indicating the use of HPC for data transmission. Based on the Lyapunov drift plus penalty function framework \cite{Lyapunov,Neely1,Neely2}, we can define a simple rule to make HPC leasing decisions. According to this rule, in every time slot, the BS should compute a threshold denoted by $Y^*=\frac{V c(t)}{2}$, where $V>0$ is the HPC cost weighting factor that can be adjusted and allows the tradeoff between queuing delay and spectrum cost. The value of $D_k(t)$ should be set to 1, if $Y_k(t) > Y^*$; otherwise it should be set to 0. If all $K$ values of $D_k(t)$ are 0, the BS should not purchase HPC. On the other hand, if any value of $D_k(t)$ is equal to 1, a purchase decision should be made.  

According to this rule, BS only requires the instantaneous values of the actual and virtual queue lengths and the instantaneous leasing cost in order to make HPC purchase decisions. The intuition behind the HPC purchase policy is also very simple. The BS makes the purchase decision whenever either or both of the following conditions hold:
\begin{enumerate}
\item The cost of HPC leasing is low.
\item The queue lengths are high, which represents higher queuing delays.
\end{enumerate}
It is important to note that the HPC cost weighting factor $V$ can be used by the BS to influence and control the HPC leasing decisions. For example, for a given value of $c(t)$, setting a lower value of $V$ will decrease the threshold value $Y^*$, and hence will result in more frequent HPC leasing. The value of $V$ can be optimized depending on the application requirements. A higher value of $V$, on the other hand, would result in less HPC deployment. As described in \cite{Neely1,Neely2}, this dynamic Lyapunov based scheme also upper bounds the queuing delay of packets at the data concentrators.

\section{Dynamic-CD-QUALITY Scheme: Mechanism to tradeoff HPC leasing cost, transmission delays and data quality}
\label{sec:Qual}
Dynamic-CD-LYAPUNOV scheme developed in the previous section allows the CR operator to tradeoff HPC leasing cost and queuing delays. However, in certain smart grid applications, the size of some data units can also be reduced in order to avoid HPC purchase decisions. For example, the least significant bits of quantized data readings provided by the smart meter, energy monitoring devices, or other sensors could be dropped. In particular for smart metering applications, since the meter readings are accumulative in nature, dropping some least significant bits in some packets will not impact on the final billing. In order to avoid HPC purchase, now the CR operator can also reduce data quality. In this section, we discuss a mechanism to tradeoff HPC leasing cost, transmission delays and the data quality.

\subsection{System Model and Problem Formulation}
In the previous section, we used the concept of delay aware virtual queues to keep track of the queuing delays and then developed a real-time solution using Lyapunov drift plus penalty function framework. This framework provides a worst case bound on the transmission delay and incorporating data quality constraint in the queuing based model is not straightforward. We, therefore, adopt a more direct problem formulation approach and we also consider hard constraints on the transmission delays and data quality. In this problem, we consider that each data concentrator can make independent HPC purchase decisions. 
 
We define a data unit, which could be a single packet or a group of packets depending on the mean packet arrival rate. In terms of data units, we assume that one data unit arrives every time slot. This arrival process is similar to the queuing based model, where we assumed that in every time slot data packets arrive with a certain mean packet arrival rate. In order to simultaneously consider transmission delay and data quality, we consider the transmission of a total of $N$ equal size data units that arrive in a certain time duration. Due to the causality constraint, data units cannot be transmitted before their arrival. We ensure this constraint by assuming that the data concentrator can only transmit a single data unit in one time slot either using free or the leased HPC spectrum. This constraint also upper bounds the maximum departure rate to 1 data unit per data concentrator. With this constraint, $N$ data units can always be transmitted in $t \geq N$ time slots. 

To complete the problem development, we introduce a hard transmission delay constraint such that the overall data units get transmitted in $T \geq N$ time slots. We also introduce data quality constraint by assuming that the size of any $M < N$ data units can be reduced. We assume that the impact of size reduction in $M$ data units on the overall data quality is acceptable if the remaining $N-M$ data units of full size gets transmitted. For simplicity, we further assume that only one size reduction is possible. In other words, all the resulting reduced size data units will always be of equal size. We also assume different HPC purchase prices for the transmission of full size and reduced size data units. However, the cost to transmit a full size data unit in any time slot will always be higher than the transmission cost of a reduced size data unit in the same time slot. The transmission cost also depends on the number of packets in each data unit. The overall objective at each data concentrator is to minimize the total cost of transmission of $N$ data units in $T$ time slots with an additional option of reducing the size of at most $M$ data units. 

\subsection{Real-time Dynamic-CD-QUALITY Scheme}
In a dynamic environment with time varying spectrum availability and HPC leasing prices, obtaining a real-time optimal solution for this problem is quite challenging. We developed a sub-optimal heuristic for this problem in \cite{icc_paper}, which we term here as our dynamic-CD-QUALITY scheme. This algorithm defines two parametrized average prices (PAP), i.e., PAP-full and PAP-reduced, respectively for the full size and the reduced size data units. The PAPs depend on the HPC purchase prices in the all the previous time slots as well as an externally controllable parameter $\beta_c$, where, $0 \leq \beta_c \leq 1$. If free TVWS spectrum is not available and the current HPC leasing price is less than PAPs then depending on some simple rules, decisions are made to purchase either a full size or a reduced size data unit. As the future TVWS spectrum availability and HPC lease prices are assumed unknown, the intuition behind the developed strategy is based on finding the lowest priced time slots by utilizing the previous information. We skip the details of this algorithm and refer interested readers to \cite{icc_paper}.

\section{Simulation Results}
\label{sec:sims}
In this section, we analyze the performance of our proposed Dynamic-CD-LYAPUNOV and dynamic-CD-QUALITY schemes that are developed for the futuristic scenario of TVWS spectrum congestion through simulations. We first explain our simulation setup followed by the numerical results.
\subsection{Simulation Setup}
We consider a NAN comprising of 60 buildings spread in a circular region of 500m radius. Depending on spectrum congestion and wireless channel conditions, the availability of free TVWS spectrum is divided into three states corresponding to no transmission, reduced size data unit transmission and full size data unit transmission. These states are generated as uniformly distributed random integers. We assume that HPCs can always be purchased by the BS or the data concentrator whenever required.  

Foreseeing that the communication network architecture employing TVWS could support several smart grid applications, we simulate different packet arrival rates. The size of data unit is accordingly determined by the mean packet arrival rate. In each time slot, we model the HPC leasing cost as a uniformly distributed random variable between $[0.1,1]$ dollar cent to transmit one full size packet per time slot. We assume that HPC leasing requirements linearly change with the corresponding increase or decrease in the number and size of the packets as well as the time slot duration\footnote{The achieved data rates will depend on packet size and the duration of time slot, which can be arbitrarily selected and any reasonable values can be assumed. Similarly, the choice of HPC leasing cost is also arbitrary and considering different values will only scale the numerical results.}. For various comparisons and graphs, total simulation duration is assumed to be 10,000 time slots.

\subsection{Numerical Results}
First, we evaluate the performance of our Dynamic-CD-LYAPUNOV scheme developed in Section~\ref{sec:Qos} and determine the role of HPC cost weighting factor $V$ to tradeoff HPC leasing cost and queuing delays. Note that in this technique, we cannot perform a tradeoff with data quality. We compare the performance of our Dynamic-CD-LYAPUNOV scheme with a static leasing scheme. In the static leasing scheme, HPC purchase decisions are made at fixed and regular intervals. In our simulations, we consider two static leasing schemes, called static scheme 1 and static scheme 2. In both the static schemes, HPC purchase decisions are made after every 1000 times slots, i.e., at $t=1000, 2000, \ldots, 10,000$. In static scheme 1, BS continuously purchases HPCs for 200 consecutive times slots (e.g., in every time slot $t=1001$ till $t=1200$), while in static scheme 2, BS continuously purchases HPCs for 150 consecutive times slots (e.g., in every time slot $t=1001$ till $t=1150$).

\begin{figure}[t]
\centering
\includegraphics[width=\columnwidth]{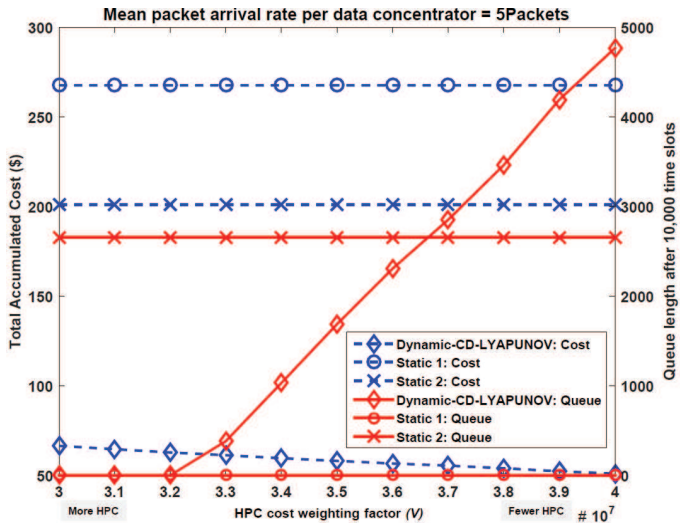}
 \caption{Total accumulated cost (\$) and queue length after $10,000$ time slots for different values of HPC cost weighting factor $V$.}
 \label{fig:figs66}
\end{figure}

Figure \ref{fig:figs66} shows the total average accumulated cost and queue length per data concentrator of our proposed Dynamic-CD-LYAPUNOV scheme for different values of $V$ at the end of 10,000 time slots. The mean packet arrival rate is assumed to be 5packets per data concentrator per time slot. We can observe that as the value of $V$ increases, the total accumulated cost decreases. We can also observe the existence of an optimal value of $V$ at which the total accumulated cost is minimum (62.77\$) and the queue length after 10,000 time slots is zero. This value turns out to be $V^*=3.2 \times 10^7$. At other values of $V$ that are lower than $V^*$, frequent and unnecessary HPC purchase decisions are made, which results in a higher cost. In these situations, due to sufficient availability of free TVWS spectrum, there is no impact on the resulting queue length. On the other hand, if the value of $V$ is increased beyond $V^*$, the total cost decreases at the expense of higher queue lengths. In this scenario, the BS can effectively make a trade-off between HPC leasing cost and queuing delay (using Little's law, we can also determine the average queuing delay as a ratio of average queue length and mean packet arrival rate). It is also obvious that static schemes do not depend on $V$. However, in the static schemes, increasing or decreasing the total number of consecutive buying time slots can be used to tradeoff HPC leasing cost and queuing delays. For example, static scheme 1 (consecutive buying for 200 times slots) results in a higher accumulated cost but the resulting queue length is also lower. On the other hand, static scheme 2 (consecutive buying for 150 times slots) results in a lower accumulated cost with a relatively higher queue length. Unlike the static schemes, our proposed real-time Dynamic-CD-LYAPUNOV scheme can adapt according to the channel variations, TVWS spectrum availability and HPC leasing cost variations and thus results in relatively lower accumulated cost and lower queue lengths. For higher values of mean packet arrival rate, we also observed similar trends (the figures are omitted due to space limitations and repeatability). 

\begin{figure}[t]
\centering
\includegraphics[width=\columnwidth]{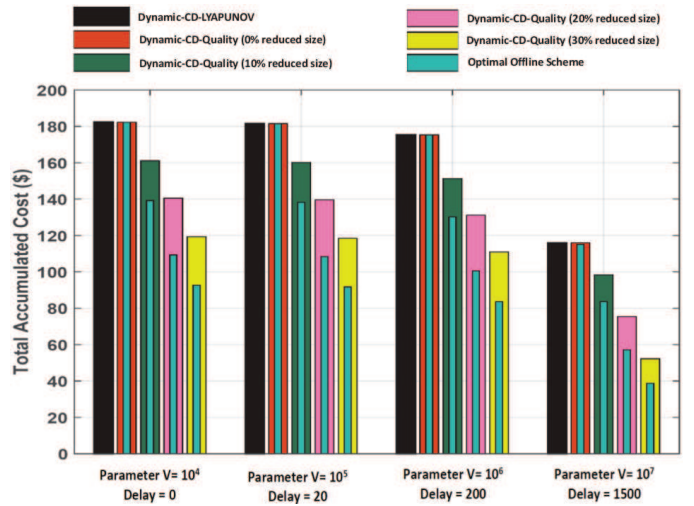}
 \caption{Total accumulated cost (\$) for different values of delay and data quality constraints for Dynamic-CD-LYAPUNOV and dynamic-CD-QUALITY schemes in comparison with optimal offline scheme. Mean packet arrival rate per data concentrator is 5packets per time slot.}
 \label{fig:figsn1}
\end{figure}

Next, we compare the performance of our Dynamic-CD-LYAPUNOV and dynamic-CD-QUALITY schemes. In the dynamic-CD-QUALITY scheme we can additionally tradeoff data quality. In our dynamic-CD-QUALITY scheme, we assumed that we can only transmit one data unit in one time slot. Therefore, for fair comparison, the maximum departure rate from the queue of any data concentrator in the Dynamic-CD-LYAPUNOV scheme is also assumed to be equal to the mean packet arrival rate (size of data unit is equal to the mean packet arrival rate). For these simulations, the mean packet arrival rate is assumed to be 5 packets per data concentrator. For the Dynamic-CD-LYAPUNOV scheme, we use Little's law to determine the average delay as a ratio of mean queue length and the mean packet arrival rate for different values of HPC cost weighting factor $V$. The average delay is then used as a delay constraint in our dynamic-CD-QUALITY scheme. For each value of delay constraint, we also change the data quality constraint as a percentage of total data units which can be transmitted with reduced quality. We consider four different data quality constraints, i.e., 0\%, 10\%, 20\% and 30\% of total $N=10,0000$ data units can be transmitted with reduced quality. We have also compared the performance of our proposed schemes with an optimal offline scheme. The optimal offline scheme assumes complete knowledge of future channel states and HPC prices. It is obvious that the performance of our real time algorithms (which do not assume any future information) cannot exceed the performance of the optimal offline algorithm. The comparison results are presented in Figure \ref{fig:figsn1}. Please note that in this figure, the thinner bars inside the dynamic-CD-QUALITY scheme represent the performance of optimal offline scheme for the same values of delay and data quality constraints. From this Figure, we can observe that for every value of delay constraint, the total accumulated cost obtained by our dynamic-CD-QUALITY scheme with 0\% data unit size reduction is almost equal to the total accumulated cost obtained by the Dynamic-CD-LYAPUNOV scheme. This demonstrates that the performance of the heuristic algorithm developed for the dynamic-CD-QUALITY scheme matches that of the Dynamic-CD-LYAPUNOV scheme. Further, as we increase the data quality constraint, i.e., the percentage of data units whose size can be reduced, the total accumulated cost also decreases. It is also interesting to observe that for any value of delay constraint, increasing the data quality constraint from 0\% to 10\% does not make  significant impact on the total accumulated cost. However, for higher values of data quality constraint, e.g., 30\% we see very huge gains in terms of overall cost reduction. Moreover, for higher values of both the delay constraint and the data quality constraint, the total accumulated cost significantly decreases for our real time schemes. It is also important to note that the performance of our real time schemes in comparison to the optimal offline scheme could be further improved particularly for the cases when there is greater flexibility in delay and data quality. The CR operator can use these tradeoff mechanisms to its advantage depending on the underlying smart grid application requirements.

\section{Conclusion}
\label{sec:conc}
In this paper, we proposed a communication network architecture that can exploit TVWS spectrum to support novel smart grid applications in large scale residential, commercial and industrial setups. Without any competition for free TVWS spectrum among the CR operators, very high data rates (several Mbps) can be achieved. We then considered the problem of QoS guarantees for smart grid applications when numerous CR operators compete for TVWS spectrum share. Exploiting the idea of HPCs that could be leased for short term usage by the CR operators, we proposed real-time Dynamic-CD-LYAPUNOV and dynamic-CD-QUALITY schemes. The developed schemes are very simple rules and algorithms that can be easily utilized by the CR operators to maintain their economic viability, while providing QoS guarantees according to smart grid application needs. Simulation results revealed that for smart grid applications, huge gains in terms of overall cost reduction could be obtained by exploiting the flexibility in data quality and delay.


\end{document}